%% file: main.tex
\documentclass[11pt,twoside]{scrartcl}
\KOMAoptions{paper=a4,fontsize=11pt,pagesize=pdftex,headinclude,BCOR=8.5mm,DIV=15}

\include{header}

\begin{document}

\include{titlepage}

\clearpage

%\tableofcontents

%\clearpage

\section*{Hintergrund}
\subsection*{Das Apollo-Programm}
Das Apollo-Programm der NASA (National Aeronautics and Space Administration, USA) hatte das Ziel, innerhalb eines Jahrzehnts Menschen auf den Mond und wieder zurück zu bringen. Dieses Projekt war ein Resultat des Wettstreits der beiden Weltmächte, den USA und der UdSSR. Die Sowjetunion hatte mit Sputnik\endnote{\cyrtext{Спутник}, russisch: Begleiter, Satellit}~1, dem ersten künstlichen Satelliten im Jahre 1957, sowie mit Juri Gagarin, der als erster Mensch 1961 ins All flog, einige bahnbrechende Erfolge in der Raumfahrt. Als Antwort darauf etablierten auch die USA ihr eigenes Weltraumprogramm. Wegen des offensichtlichen Vorsprungs der UdSSR in der erdnahen Raumfahrt strebten die USA schließlich die Landung von Menschen auf dem Mond an, um noch Aussichten zu haben, die Sowjetunion in der Raumfahrt zu überholen \autocite{dupas_gab_1994}.

\begin{figure}[!ht]
\centering
\resizebox{0.73\hsize}{!}{\includegraphics{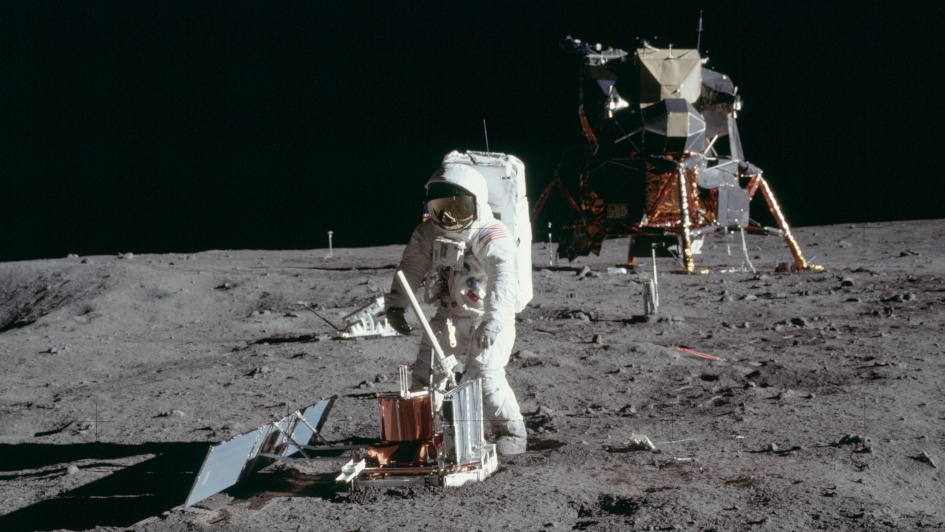}}
\caption{Die ersten Menschen landeten während der Apollo~11-Mission auf dem Mond. In diesem Bild nimmt Edwin Aldrin ein Seismometer auf dem Mond in Betrieb. Im Hintergrund steht das Landemodul (LM, Lunar Module) Eagle (Bild: NASA).}
\label{f:apollo11_aldrin}
\end{figure}

Von den 14 Flügen wurden drei (Apollo~4 bis 6) als Test ohne Besatzung durchgeführt. Neun von elf bemannten Apollo\endnote{Der Name "`Apollo"' wurde vom damaligen Leiter des NASA-Raumfahrtprogramms, Abe Silverstein nach dem Gott Apoll gewählt, der nach der griechisch-römischen Mythologie den Sonnenwagen zieht \parencite{sapienza_nasa_2009}.}-Missionen flogen zum Mond, wobei sechs auf dem Mond landeten. Die erste Landung erfolgte mit Apollo~11 am 20.~Juli 1969 (Abb.~\ref{f:apollo11_aldrin}); mit Apollo~17 erkundete die vorerst letzte Besatzung im Dezember 1972 unseren Trabanten.

\subsection*{Apollo 11}

Die Mission Apollo~11 startete am 16.~Juli 1969 mit einer Saturn~V-Rakete\endnote{siehe Material "`Wie kamen die Astronauten von Apollo 11 zum Mond?"', Raum für Bildung, \url{http://www.haus-der-astronomie.de/raum-fuer-bildung}} vom Kennedy Space Center in Florida \autocite[][S.~92]{orloff_apollo_2000}. Sie setzte sich aus der Kommandoeinheit (Command and Service Module, CSM), genannt Columbia, sowie der Landeeinheit (Lunar Module, LM), genannt Eagle zusammen. Die Besatzung bestand aus dem Kommandanten Neil Armstrong, dem Piloten der Landefähre Edwin Aldrin, sowie dem Piloten der Kommandoeinheit Michael Collins. Nach dem Abkoppeln des LM vom CSM landeten Armstrong und Aldrin am 20.~Juli 1969 auf dem Mond \autocite[][S.~94]{orloff_apollo_2000}. Collins blieb wären der gesamten Mission an Bord des CSM. Armstrong betrat nach etwa 7 Stunden am 21.~Juli 1969 als erster Mensch den Mond. Aldrin folgte ihm wenige Minuten später. Nach etwa $21~1/2$ Stunden traten die beiden Astronauten die Rückkehr zum CSM an. Die Landung auf der Erde erfolgte am 24.~Juli 1969 durch eine geplante Wasserung im Pazifik \autocite[][S.~98]{orloff_apollo_2000}.

\begin{figure}[!ht]
\centering
\resizebox{0.75\hsize}{!}{\includegraphics{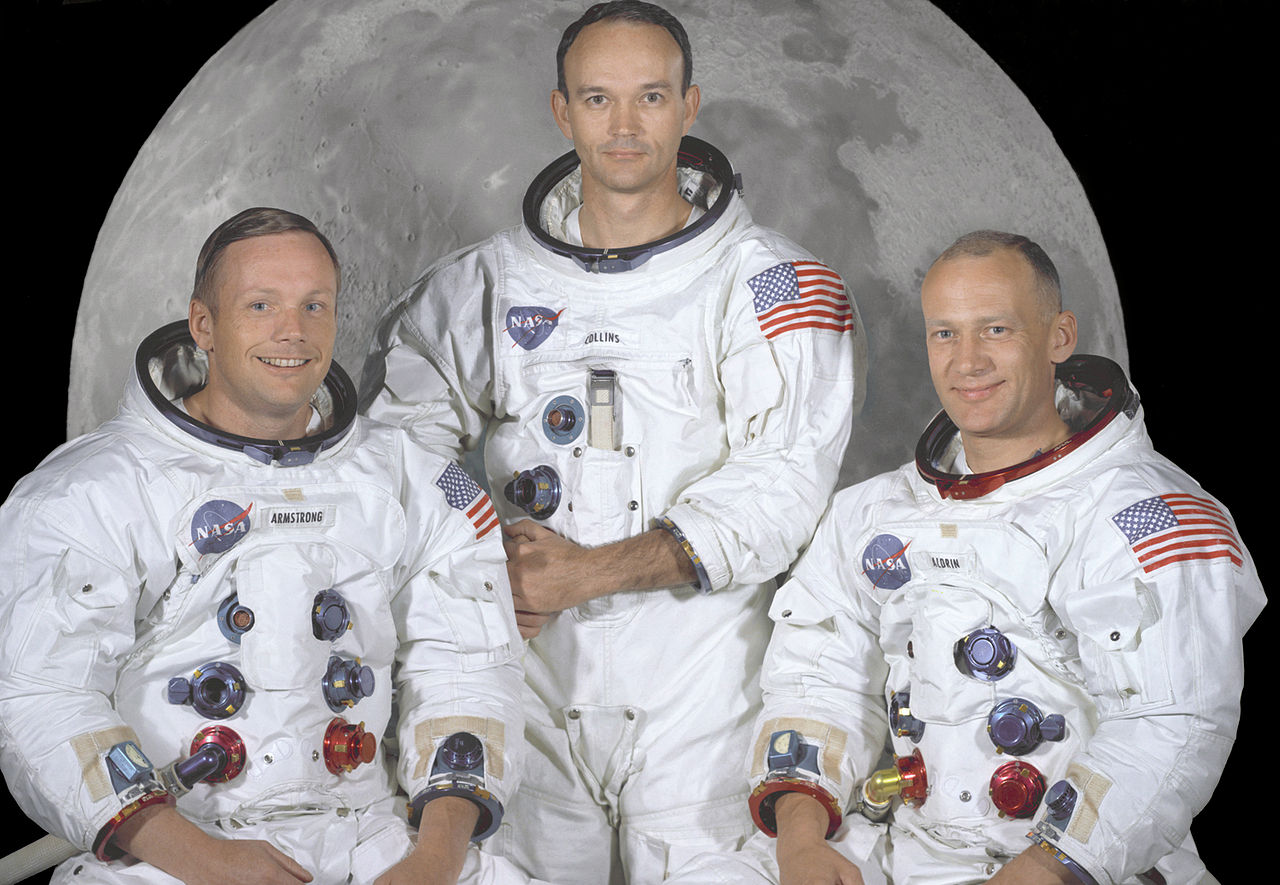}}
\caption{Die Besatzung der Apollo~11-Mission. Von links nach rechts: Neil Armstrong, Michael Collins, Edwin "`Buzz"' Aldrin (Bild: NASA)}
\label{f:apollo11_crew}
\end{figure}

\subsection*{Bestimmung der Mondentfernung}

\begin{figure}[!ht]
\centering
\resizebox{0.75\hsize}{!}{\includegraphics{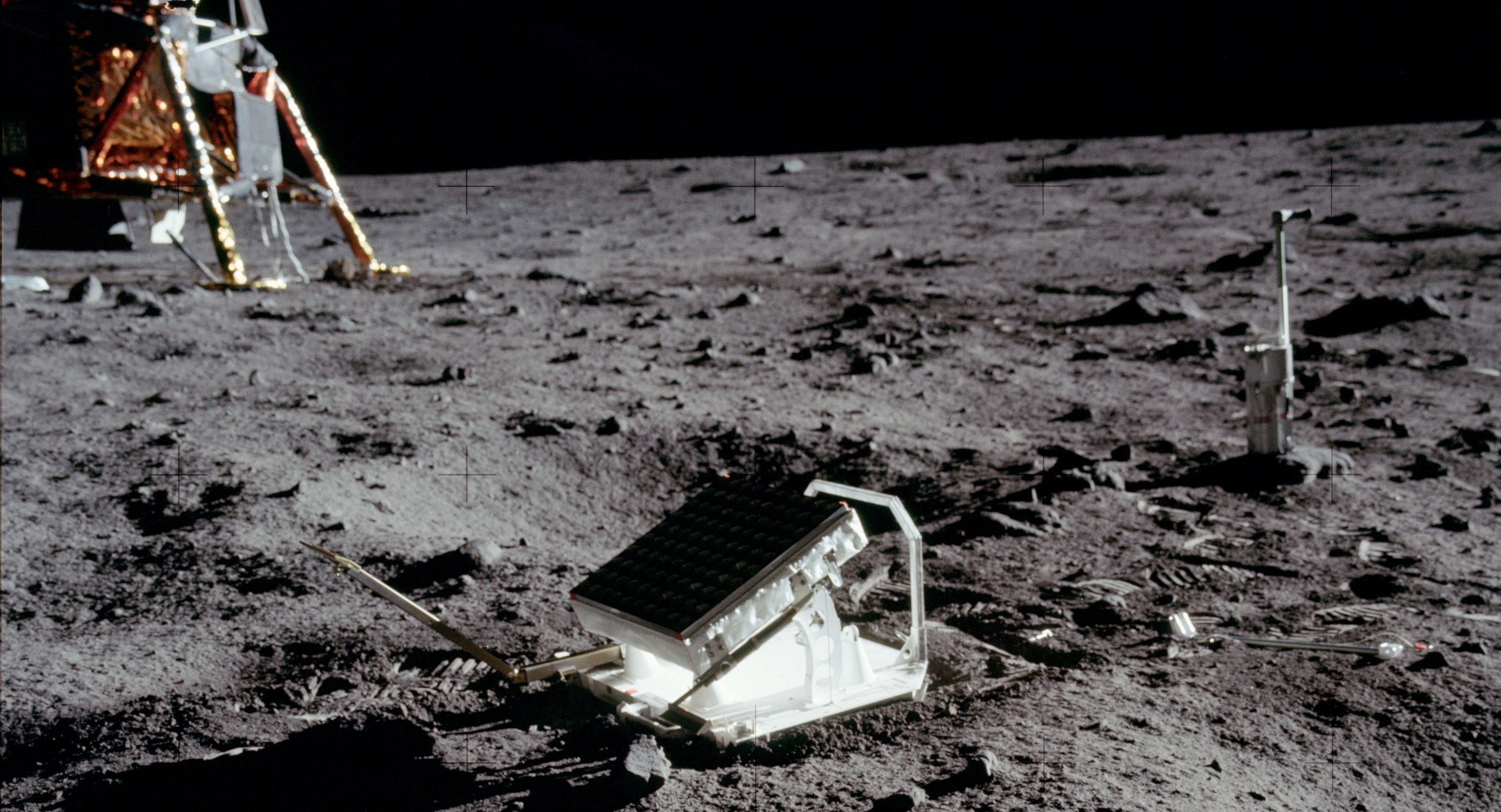}}
\caption{Mit dem Reflektor lässt sich seit der Landung von Apollo~11 die Entfernung des Mondes exakt durch Lunar Laser Ranging bestimmen (Bild: NASA).}
\label{f:apollo11_reflektor}
\end{figure}

Eines der Experimente, das mit Apollo~11 auf dem Mond aufgestellt wurde, ist der Reflektor (Abb.~\ref{f:apollo11_reflektor}) mit dem man die Distanz zum Mond bestimmen kann \parencite{soffel_lasermessungen_1997}. Hierbei wird die Laufzeit eines von der Erde ausgesandten Laserpulses gemessen, der von diesem Reflektor zur Erde zurück geworfen wird. Über die Lichtgeschwindigkeit erfolgt eine sehr genaue Bestimmung der Mondentfernung. Diese Methode nennt man "`Ranging"'. Bis dahin nutzte man eine ähnliche Technik, bei der mit Radar (Radio Detection and Ranging) ausgesandte Radiowellen von der Mondoberfläche reflektiert werden \parencite{yaplee_radar_1958,yaplee_lunar_1959}. Beide Methoden sind jedoch technisch sehr aufwendig und benötigen spezielle Apparaturen sowie eine ausgefeilte Analysetechnik.

Im Prinzip lässt sich jedoch jede Art von elektromagnetischen Signalen für eine Entfernungsbestimmung nutzen. Eine für den Schulunterricht recht spannende Variante ist die Analyse der Funksignale, die während der Apollo-Missionen zwischen der Besatzung und der Missionskontrolle am Erdboden ausgetauscht wurden. In manchen Fällen hört man das Echo des Signals eines Gesprächpartners, das durch die andere Seite über das Mikrofon wieder zum Ausgangspunkt zurück gesendet wird. Die Dauer zwischen der Originalbotschaft und dem Echo entspricht der doppelten Signallaufzeit zwischen Erde und Mond.

\begin{figure}[!ht]
\centering
\resizebox{\hsize}{!}{\includegraphics{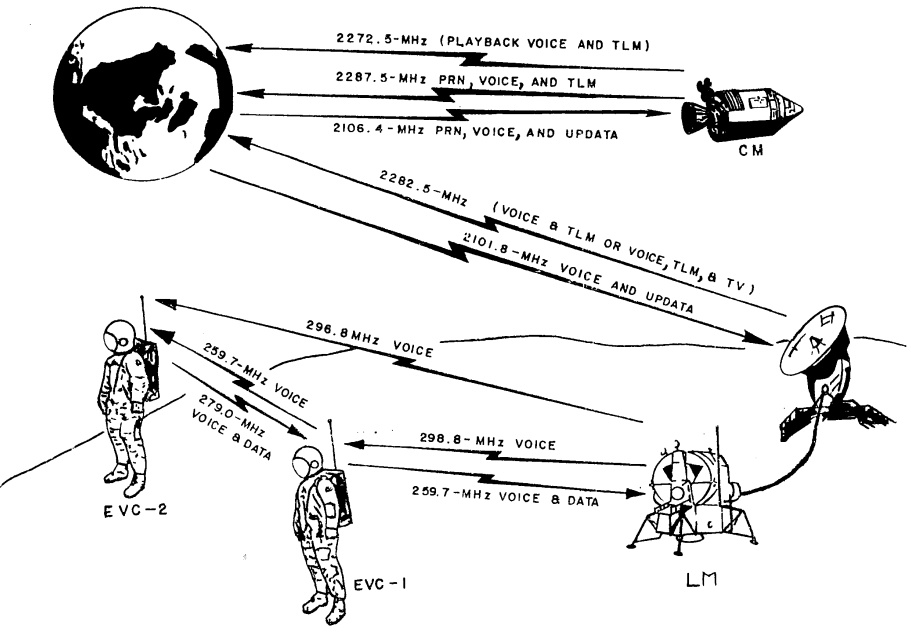}}
\caption{Übersicht der Funkverbindungen zwischen Apollo~11 und der Erde (Bild: NASA, \textcite{corliss_histories_1974}).}
\label{f:apollo11_funk1}
\end{figure}

\subsection*{Datenverkehr zwischen Apollo 11 und Erde}

\begin{figure}[!ht]
\centering
\resizebox{\hsize}{!}{\includegraphics{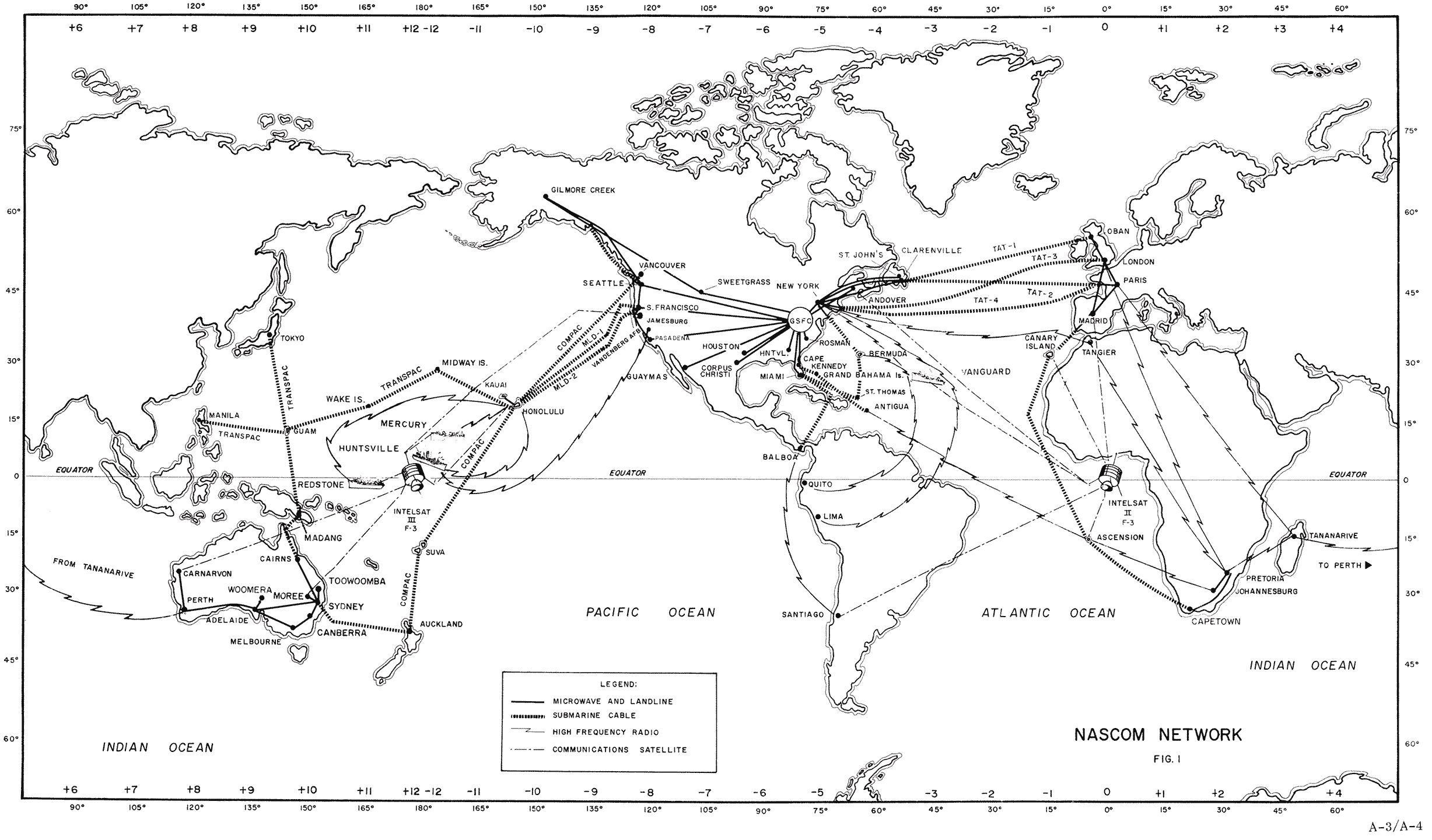}}
\caption{Übersichtskarte des Netzwerks von Datenverbindungen auf der Erde während der Apollo~11-Mission (Bild: NASA, \textcite{graham_manned_1970}).}
\label{f:apollo11_funk2}
\end{figure}

Während einer Apollo-Mission wurden Daten zwischen den Raumfahrzeugen und der Bodenkontrolle ausgetauscht. Sensoren schickten Informationen zum technischen Zustand der Ausrüstung sowie der Körperfunktionen der Astronauten zur Erde. Zudem tauschten sich die Astronauten mit der Missionskontrolle per Funkverkehr aus. Weiterhin wurden Fernsehbilder live von der Mondoberfläche zur Erde übermittelt. Dies alles benötigte eine komplexe Struktur aus verschiedenen Systemen zur Datenübertragung, die allmählich seit den ersten Raumflügen etabliert wurde. Der generelle Aufbau des Netzwerks für den Datenverkehr während der Apollo~11-Mission ist in Abb.~\ref{f:apollo11_funk2} dargestellt.

\medskip
Als Empfangs- und Relaisstationen dienten Radioantennen, Satelliten, Flugzeuge und Schiffe, die alle per Kabel- und Funkstrecken sowie Verteilerknoten miteinander verbunden waren. Dies führte neben der Signallaufzeit durch die Entfernung zwischen Erde und Mond zu einer zusätzlichen Verzögerung im Funk- und Datenverkehr.

\subsection*{Signallaufzeit und Lichtgeschwindigkeit}
Elektromagnetische Wellen, wie auch Funk- und Radiowellen, breiten sich im Vakuum stets und in jedem Bezugssystem mit der konstanten Geschwindigkeit $c = \unit[299792458]{m/s}$, der Vakuumlichtgeschwindigkeit aus. Daher lässt sich aus der Laufzeit $\Delta t$ eines Funksignals im Vakuum die zurückgelegte Strecke $\Delta x$ ermitteln (Abb.~\ref{f:apollo11_funk1}).

\begin{equation}
\label{e:entfernung}
    \Delta x = c \cdot \Delta t
\end{equation}

\subsection*{Unsicherheit der Messung und Statistik}
\label{s:stat}
Jede Messung ist geprägt von Unsicherheiten bei der Bestimmung des Messwerts. Diese Unsicherheiten können statistischer oder auch systematischer Natur sein. In der vorliegenden Aufgabe besteht die systematische Unsicherheit darin, dass die Dauer zwischen dem Ausgangssignal und seinem Echo nicht nur von der Signallaufzeit zwischen Mond und Erde bestimmt wird. Zusätzliche Verzögerungen wurden durch die Laufzeiten der Funk- und Kabelstrecken sowie der Signalverarbeitung auf der Erde hervorgerufen. Auch wenn man die Zeit zwischen Signal und Echo exakt messen könnte, ist das Ergebnis systematisch um einen bestimmten Wert verschoben. Dieser Einfluss wird in der vorliegenden Aufgabe zunächst vernachlässigt. Seine Größenordnung wird am Ende der Aufgabe bestimmt.

\medskip
Statistische Unsicherheiten entspringen der Ungenauigkeit des Messprozesses selbst. Kein Wert kann exakt abgelesen oder bestimmt werden. Dadurch kommt es zu einer Streuung der Messwerte. Um dem wahren Wert $x$ möglichst nahe zu kommen, bedient man sich statistischer Verfahren, die auf einer möglichst großen Anzahl von Messungen beruhen, die im Prinzip alle denselben Wert ergeben müssten, wäre der Messprozess perfekt präzise. Eine Messreihe ist daher eine Stichprobe aller möglichen (unendlich vielen) Messungen. Geht man von einer zufälligen Verteilung der statistischen Unsicherheit der Einzelmessungen aus, sollten alle Werte gleichmäßig um den wahren Wert $x$ streuen. Das Maß der Streuung ermöglicht eine Angabe über die Genauigkeit des ermittelten Werts.

\medskip
Um sich dem wahren Wert einer Messreihe anzunähern, bietet sich die Bestimmung des {\em arithmetischen Mittels} $\bar{x}$ an. $\bar{x}$ ist der Wert den man erhielte, wenn $\bar{x}$ mit der Anzahl der Messwerte $n$ multipliziert denselben Wert ergäbe, wie wenn man alle Messwerte $x_i$ aufaddiert.

\bigskip
\begin{eqnarray}
    \nonumber
    n\cdot\bar{x} &=& x_1+x_2+\ldots +x_n = \sum_{i=1}^n x_i \\[5pt]
    \label{e:mittelwert}
    \Leftrightarrow \bar{x} &=& \frac{x_1+x_2+\ldots +x_n}{n} = \frac{1}{n}\sum_{i=1}^n x_i
\end{eqnarray}

%\bigskip
%Ein wesentliches Merkmal des arithmetischen Mittels ist, dass die Summe aller $x_i-\bar{x}$ exakt Null ergibt.
%
%\bigskip
%\begin{equation*}
%    \sum_{i=1}^n \left(x_i-\bar{x}\right) = \sum_{i=1}^n x_i - \sum_{i=1}^n \bar{x} = \sum_{i=1}^n x_i - n\cdot \bar{x} = n\cdot \bar{x} - n\cdot \bar{x} = 0
%\end{equation*}

\bigskip
Eine Größe, die eine Aussage über die Qualität der Messreihe (Stichprobe) macht, ist die {\em Standardabweichung} $s$. Sie beschreibt die Streuung der Messwerte um den Mittelwert.

\bigskip
\begin{equation}
\label{e:stdev}
    s = \sqrt{\frac{1}{n-1}\sum_{i=1}^n \left(x_i-\bar{x}\right)^2}
\end{equation}

\bigskip
Allerdings ist der Mittelwert $\bar{x}$ einer Stichprobe nur eine Näherung für den wahren Wert $x$. Theoretisch gleicht der Mittelwert dem wahren Wert für eine unendliche Anzahl von Messungen. Inwieweit das arithmetische Mittel dem wahren Wert entspricht, wird durch die {\em Standardabweichung des Mittelwerts} $SEM$ (auch Standardfehler des Mittelwerts) ausgedrückt. Die $SEM$ gibt an, wie sehr alle möglichen Mittelwerte unendlich vieler Stichproben um den wahren Wert streuen.

\bigskip
\begin{equation}
\label{e:sem}
    SEM = \frac{s}{\sqrt{n}} = \sqrt{\frac{1}{n\cdot(n-1)}\sum_{i=1}^n \left(x_i-\bar{x}\right)^2}
\end{equation}

\bigskip
Insbesondere gilt:

\bigskip
\begin{equation}
    \lim_{n\to\infty}{SEM} = \lim_{n\to\infty}{\frac{s}{\sqrt{n}}} = 0
\end{equation}

\bigskip
Das entspricht der vorherigen Aussage, dass für unendlich viele Messungen der Mittelwert dem wahren Wert entspricht. Somit lässt sich für den wahren Wert $x$ schreiben:

\bigskip
\begin{equation}
\label{e:erwartungswert}
   x = \bar{x} \pm SEM 
\end{equation}

%%%%%%%%%%%%%%%%%%%%%%%%%%%%%%%%%%%%%%%%%%%%%%%%%%%%%%%%%

\clearpage
\section*{Vorbereitung der Aktivität}

\subsection*{Hinweise für Lehrpersonen}
Lesen Sie das Kapitel mit den Hintergrundinformationen sorgfältig. Zusätzliche Literatur finden Sie am Ende dieses Dokuments.

\medskip
Machen Sie sich mit den Aufgaben der Schülerinnen und Schüler vertraut. Fertigen Sie ausreichend Kopien der Arbeitsblätter an. Für die Aktivität benötigen sie Taschenrechner.

\medskip
Für die Auswertung werden zwingend Computer benötigt. Tablets sind ungeeignet. Die Software {\Aud} ist kostenlos für die Betriebssysteme Windows, MacOS und Linux erhältlich:

\bigskip
\url{https://www.audacity.de}

\bigskip
Die Nutzung ist intuitiv und schnell zu erlernen. Es ist jedoch ratsam, dass Sie vorab die Auswertung trainieren. Die Benutzeroberfläche ist in Abb.~\ref{f:audacity1} dargestellt. Eine Anleitung finden Sie in der Beschreibung der Aufgabe.

\medskip
Zusammengefasst wird eine Audiodatei eingeladen und mit der Software grafisch zeitaufgelöst dargestellt. Bei der Datei handelt es sich um eine digitalisierte Version eines Audiomitschnitts während der Apollo~11-Mission, die von der NASA kostenlos zum Download angeboten wird. Diese finden Sie zum Download auf der Projektseite

\bigskip
\url{https://www.haus-der-astronomie.de/raum-fuer-bildung}

\bigskip
oder im Internet Archive:

\bigskip
\url{https://archive.org/download/Apollo11Audio/175-AAA.mp3}

\bigskip
Weitere Dateien des Funkverkehrs erhalten Sie unter:

\bigskip
\url{https://archive.org/details/Apollo11Audio}

\bigskip
Die Schülerinnen und Schüler ermitteln mit der Software {\Aud} die Zeitmarken von einzelnen Funksignalen und deren Echos.

\medskip
Teilen Sie die Schülerinnen und Schüler in Gruppen ein. Optimal wären Zweiergruppen. Die Gruppenstärke orientiert sich aber natürlich auch nach dem Verhältnis zwischen der Anzahl der vorhandenen Computer und den Lernenden. Möglichst sollten Kopfhörer bei der Auswertung der Audiodatei benutzt werden sollten, um die anderen Gruppen nicht zu beeinträchtigen. Beachten Sie, dass dadurch immer nur ein Mitglied einer Gruppe die Radiosignale hören kann. Soweit es die Zeit erlaubt, können sich die Gruppenmitglieder bei der Bedienung der Software abwechseln. Durch die Visualisierung der Töne können alle Mitglieder einer Gruppe gemeinsam die Daten analysieren.

\medskip
Für eine Teilaufgabe (Ermittlung der tatsächlichen Mondentfernung am 21.~Juli 1969) ist optional eine Internetverbindung erforderlich. Falls das nicht möglich ist, wird der Wert in der Aufgabe vorgegeben.

\clearpage
\subsection*{Thematische Einführung (Vorschlag)}

Informationen und Videos zur Einstimmung auf das Apolloprogramm helfen bei der Annäherung an das Thema.

\bigskip
DLR: Best of Apollo - 50 Jahre Mondlandung\\
\url{https://youtu.be/8iNq_S8O7K4} (3:41 min)

\bigskip
DLR\_next -- Apollo 11 -- Mond Special\\
\url{www.dlr.de/next-apollo}

\bigskip
Fragen Sie die Schülerinnen und Schüler, was Sie über die Mondlandungen und das Apollo-Programm wissen und was sie daran interessiert. Stichpunkte können an der Tafel gesammelt werden.

\bigskip
Falls Sie ausreichend Zeit haben, lassen Sie die Lernenden im Internet über Apollo, insbesondere Apollo 11 recherchieren. Die Thematik lässt sich sehr gut durch Referate zu verschiedenen Themen der bemannten Raumfahrt innerhalb eines längeren Projekts behandeln.

\bigskip
Einen guten Überblick über die verschiedenen Phasen einer Apollo-Mission ermöglichen die folgenden drei Animationsvideos. Sie wurden in englischer Sprache verfasst, beinhalten jedoch deutsche Untertitel.

\bigskip
Wie das Apollo Raumschiff funktioniert: Teile 1 - 3\\
\url{https://youtu.be/8dpkmUjJ8xU} (3:57 min, Englisch)\\
\url{https://youtu.be/tl1KPjxKVqk} (5:17 min, Englisch)\\
\url{https://youtu.be/qt_xoCXLXnI} (4:01 min, Englisch)

\bigskip
Die originale Fernsehübertragung des Starts von Apollo~11 vom 16.~Juli 1969 ist im folgenden Video zu sehen.

\bigskip
1969 Apollo 11 Saturn V launch, 1969 TV broadcast\\
\url{https://youtu.be/xdxzMPi19sU} (38:39 min, Englisch)

\bigskip
Die originalen Fernsehbilder der Apollo~11-Mission vom Mond werden von der NASA unter folgendem Link angeboten.

\bigskip
Apollo 11 HD Videos\\
\url{https://www.nasa.gov/multimedia/hd/apollo11_hdpage.html}

\bigskip
Alle Fotos, die jemals während des Apollo-Programms gemacht wurden, sind im Project Apollo Archive gesammelt und sind für jeden frei verfügbar.

\bigskip
Project Apollo Archive\\
\url{https://www.flickr.com/photos/projectapolloarchive/albums}

%%%%%%%%%%%%%%%%%%%%%%%%%%%%%%%%%%%%%%%%%%%%%%%%%%%%%%%%%%%%%

\clearpage
\section*{Aufgaben}
\subsection*{Ziel}

Viele Methoden zur Bestimmung der Entfernung des Monds sind schwierig und benötigen einen großen technischen Aufwand. In dieser Aufgabe benutzen Sie das Konzept der Messung einer Sig\-nal\-lauf\-zeit, dessen Geschwindigkeit bekannt ist.

\medskip
Während der Apollo~11-Mission tauschten die Astronauten Informationen über Funk mit der Bodenstation auf der Erde aus. Funkwellen sind elektromagnetische Wellen wie das Licht. Sie breiten sich daher mit Lichtgeschwindigkeit aus.

\medskip
An einigen Stellen der Gespräche kam es zu Echos. Gesprochene Wörter, die die Bodenkontrolle an die Crew von Apollo 11 geschickt hatte, wurden dort manchmal vom Mikrofon eines Astronauten erfasst und wieder zur Bodenstation zurück geschickt. Sie kamen also zugleich mit den Kommentaren der Astronauten auf der Erde an und erscheinen in der Aufzeichnung als Echo. Die Verzögerung zwischen der Originalnachricht und dem Echo entspricht damit der doppelten Signallaufzeit zwischen Erde und Mond.

\medskip
Alle Funkverbindungen wurden während der Mondmission aufgezeichnet und stehen heute in digitalisierter Form zur Verfügung.

\medskip
\url{https://archive.org/details/Apollo11Audio}

\medskip
Sie werden den Sprechfunk an einigen Stellen auswerten und die Verzögerung zwischen den Nachrichten und ihren Echos bestimmen. Mit der Lichtgeschwindigkeit können Sie damit die Entfernung berechnen. Die Analyse erfolgt mit der Software {\Aud}, die den Ton grafisch darstellt. Einzelne Wörter oder gar Silben können damit markiert und deren Zeiten ermittelt werden.

\subsection*{Bedienung von {\Aud}}

{\Aud} ist eine Software zum Schneiden und Bearbeiten von Audiodateien. Für die Auswertung sind aber nur einige Funktionen notwendig. Die für diese Aufgabe benötigte Audiodatei

\medskip
\url{https://archive.org/download/Apollo11Audio/175-AAA.mp3}

\begin{figure}[!ht]
\centering
\resizebox{\hsize}{!}{\includegraphics{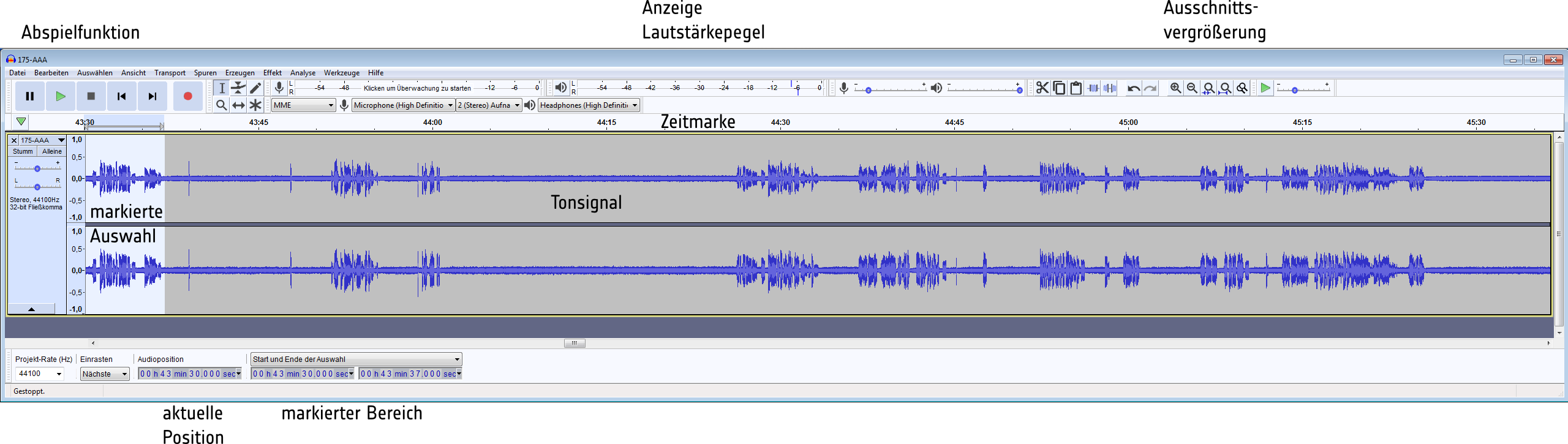}}
\caption{Benutzeroberfläche der Audioschnittsoftware {\Aud} mit Beschriftung der wichtigsten Funktionen.}
\label{f:audacity1}
\end{figure}

\medskip
wird Ihnen zur Verfügung gestellt. Sie können sie einladen, indem Sie entweder über den Menüpunkt \texttt{Datei} die Audiodatei auswählen oder einfach die Datei aus einem externen Dateimanager (z. B. Explorer bei Windows) in den Arbeitsbereich ziehen. Die Benutzeroberfläche nach dem Laden dieser Datei ist in Abb.~\ref{f:audacity1} dargestellt.

Die einzelnen Tonkanäle werden im Arbeitsbereich dargestellt. Oberhalb der Darstellung des Ton\-signals ist die Zeit ab Beginn der Datei angezeigt. Die Menüzeile enthält Schaltflächen zum Abspielen der Datei (links) und zum Hinein- und Herauszoomen (rechts). Klickt man mit der linken Maustaste in den Bereich des Tonsignals, kann die Audiodatei ab dieser Zeitmarke abgespielt werden. Diese Zeitmarke wird unten links als aktuelle Position angegeben.

\begin{figure}[!ht]
\centering
\resizebox{\hsize}{!}{\includegraphics{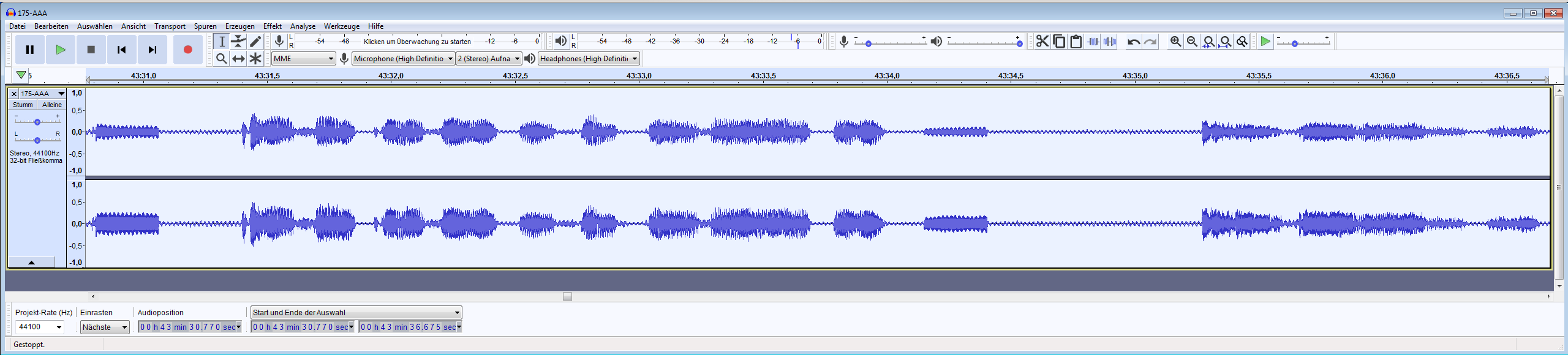}}
\caption{Vergrößerte Darstellung des Funkverkehrs.}
\label{f:audacity2}
\end{figure}

\begin{figure}[!ht]
\centering
\resizebox{\hsize}{!}{\includegraphics{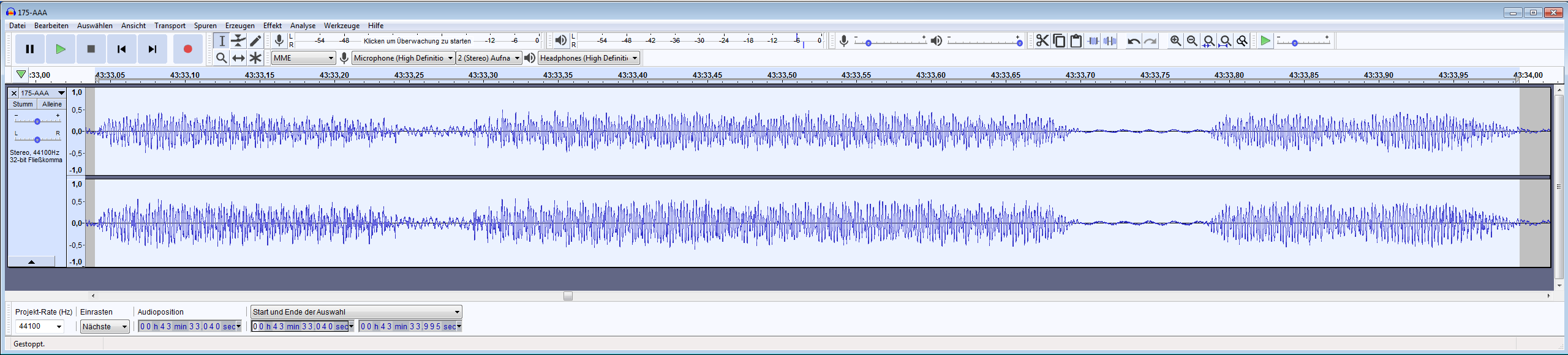}}
\caption{Vergrößerte Darstellung des Funkverkehrs.}
\label{f:audacity3}
\end{figure}

\medskip
Zusätzlich kann man mit der Computermaus durch Linksklicken und Ziehen einen Bereich markieren. Die Zeitspanne dieses Bereichs wird ebenfalls unten angezeigt. Beim Abspielen wird dann lediglich dieser Bereich wiedergegeben. Zudem lässt sich mit der Lupe mit den nach außen angedeuteten Pfeilen der Ausschnitt auf die verfügbare Ansicht vergrößern (Abb.~\ref{f:audacity2}). Das ist Hilfreich, wenn bei der Auswertung die Signale einzelner Wörter und Silben identifiziert werden müssen. Die Lupe mit den nach innen gerichteten Pfeilen stellt das gesamte Audiosignal dar. Weiteres Hineinzoomen ist jederzeit möglich (Abb.~\ref{f:audacity3}).

\bigskip

\medskip
Statt mit der Maus kann der zu markierende Bereich als Zahlenwerte für den Anfangs- und den Endzeitpunkt in der unteren Eingabemaske eingetragen werden.

\bigskip

\subsection*{Analyse der Audiodaten}
Laden Sie die Audiodatei ein und bestimmen Sie mittels {\Aud} die Zeitintervalle zwischen verschiedenen Funkechos. Textpassagen, die Funkechos enthalten, sind in Tab.~\ref{t:echolist} aufgeführt.

\sbox0{\endnotemark}
\begin{table}[!ht]
    \centering
    \caption{Liste mit Auszügen des Funkverkehrs zwischen der Bodenstation in Houston und der Apollo~11-Crew mit Echos. Die Zeitangaben der Audiodatei \texttt{175-AAA.mp3} sind in Stunden (h), Minuten (m) und Sekunden (s) ab Beginn der Datei. 0h00m00s entspricht dem 21.07.1969, 2h56m27s UT\usebox0.}
    \label{t:echolist}
    \begin{tabular}{p{11cm}cc}
    \hline
         & \multicolumn{2}{c}{Zeitmarke} \\
    Textpassage & Beginn & Ende \\
    \hline
    Okay, this looks good, Neil.                   & 0h34m25s & 0h34m30s \\
    Columbia, Columbia! This is Houston AOS. Over. & 0h43m30s & 0h43m37s \\
    {\em Längere Unterhaltung zwischen Houston und Michael Collins.} & 0h44m28s & 0h45m26s \\
    Sounded a little wet. & 1h00m05s & 1h00m10s \\
    Columbia, this is Houston. Over. & 1h01m12s & 1h01m18s \\
    Columbia, Columbia! This is Houston. Over. & 1h45m51s & 1h45m57s \\
    \hline
    \end{tabular}
\end{table}
\endnotetext{\sloppy Universal Time: weltweite Referenzzeit, die sich am Nullmeridian orientiert}

\medskip
Sie können unterschiedliche Verfahren zur Bestimmung der Signallaufzeit ausprobieren. Oft ist es möglich, den Beginn oder das Ende einer Silbe mit entsprechendem Echo zu identifizieren. An vielen Stellen können Sie sogar sowohl den Anfang und das Ende einer Passage mit Echo ausfindig machen. Markieren Sie diese Stellen so gut Sie können und notieren Sie sich die Zeiten (Zeitangabe in der Benutzeroberfläche unten).

\medskip
Danach berechnen Sie für alle Signal-Echo-Paare die Zeitdifferenzen. Dadurch erhalten Sie eine Messreihe, die für jede analysierte Stelle eine Zeitdauer enthält.

\subsection*{Bestimmung des Mittelwerts}
Wie auf Seite~\pageref{s:stat}f.~beschrieben, ist der Mittelwert eine gute Näherung für den gesuchten wahren Wert einer Messreihe. Er wird berechnet durch (Gl.~\ref{e:mittelwert}):

\begin{equation}
\nonumber
    \bar{x} = \frac{x_1+x_2+\ldots +x_n}{n} = \frac{1}{n}\sum_{i=1}^n x_i
\end{equation}

\medskip
Ermitteln Sie den Mittelwert der von Ihnen erstellten Messreihe der Signallaufzeiten entsprechend dieser Rechenvorschrift. Optional können Sie eine Software zur Tabellenkalkulation wie MS Excel benutzen. Der Mittelwert wird dort mit der Funktion \textrm{MITTELWERT} bzw. \textrm{AVERAGE} berechnet.

\subsection*{Optional: Bestimmung der Standardabweichung}

Berechnen Sie mit den Gleichungen~\ref{e:stdev} und \ref{e:sem} die Standardabweichungen des Messreihe und des Mittelwerts. Optional können Sie eine Software zur Tabellenkalkulation wie MS Excel benutzen. Die Standardabweichung der Messreihe wird dort mit der Funktion \textrm{STABW.S} bzw. \textrm{STDEV.S} berechnet. Die Länge der Messreihe wird durch die Funktion \textrm{ANZAHL} bzw. \textrm{COUNT} bestimmt.

\medskip
Schreiben Sie gemäß Gl.~\ref{e:erwartungswert} die durch Ihre Auswertung abgeleitete Signallaufzeit nieder.

\subsection*{Berechnung der Mondentfernung}

Berechnen Sie nun mit Ihren Ergebnissen mittels Gl.~\ref{e:entfernung} die Entfernung zwischen Erde und Mond. Bedenken Sie, dass die Funksignale den Weg zweimal durchlaufen. Geben Sie die Strecke in Kilometern an.

\subsection*{Option 1: Die tatsächliche Entfernung des Monds zur Erde}

Die mittlere Entfernung des Monds beträgt für den Zeitraum der Messungen:

\begin{equation}
    \unit[387932]{km}
\end{equation}

\subsection*{Option 2: Bestimmung der tatsächlichen Entfernung des Monds zur Erde}

\begin{figure}[!ht]
\centering
\resizebox{\hsize}{!}{\includegraphics{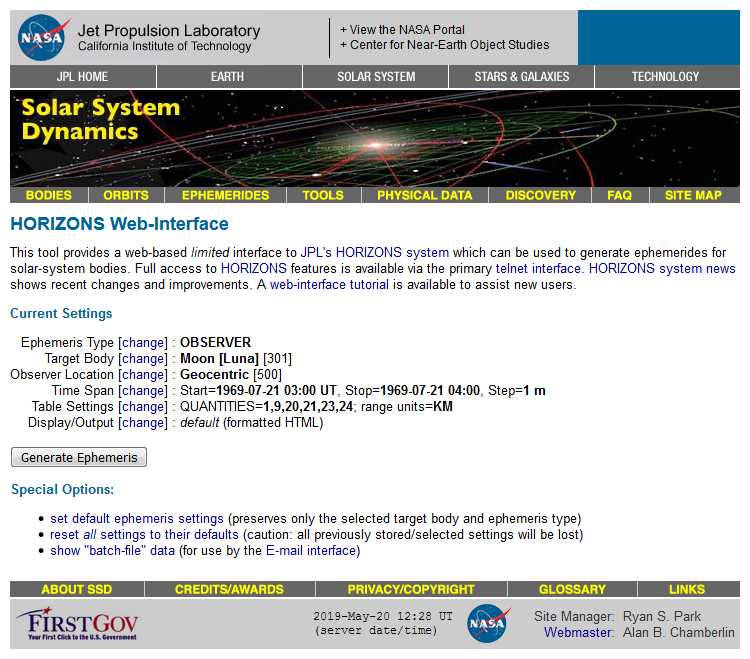}}
\caption{Benutzeroberfläche des JPL/Horizons-Webtools zur Bestimmung der Mondentfernung.}
\label{f:jpl1}
\end{figure}

Im Mittel ist der Mond etwa \unit[384.400]{km} von der Erde entfernt. Wegen der leicht elliptischen Mondbahn verändert sich der Abstand jedoch ständig. Um die Analyse der Apollo-Funksignale zu verifizieren, wird der für den entsprechenden Zeitpunkt exakte Wert benötigt. Diesen kann man sich über eine Online-Abfrage berechnen lassen. Öffnen Sie dazu die Webseite:

\bigskip
\url{https://ssd.jpl.nasa.gov/horizons.cgi}

\bigskip
Um die Entfernung des Monds zu ermitteln, müssen Sie zunächst die Ausgabeparameter (Abb.~\ref{f:jpl1}) einstellen. Das erfolgt durch einen Klick auf \texttt{\textcolor{blue}{change}} neben den aufgelisteten Kategorien.

\bigskip

\begin{figure}[!ht]
\centering
\resizebox{0.9\hsize}{!}{\includegraphics{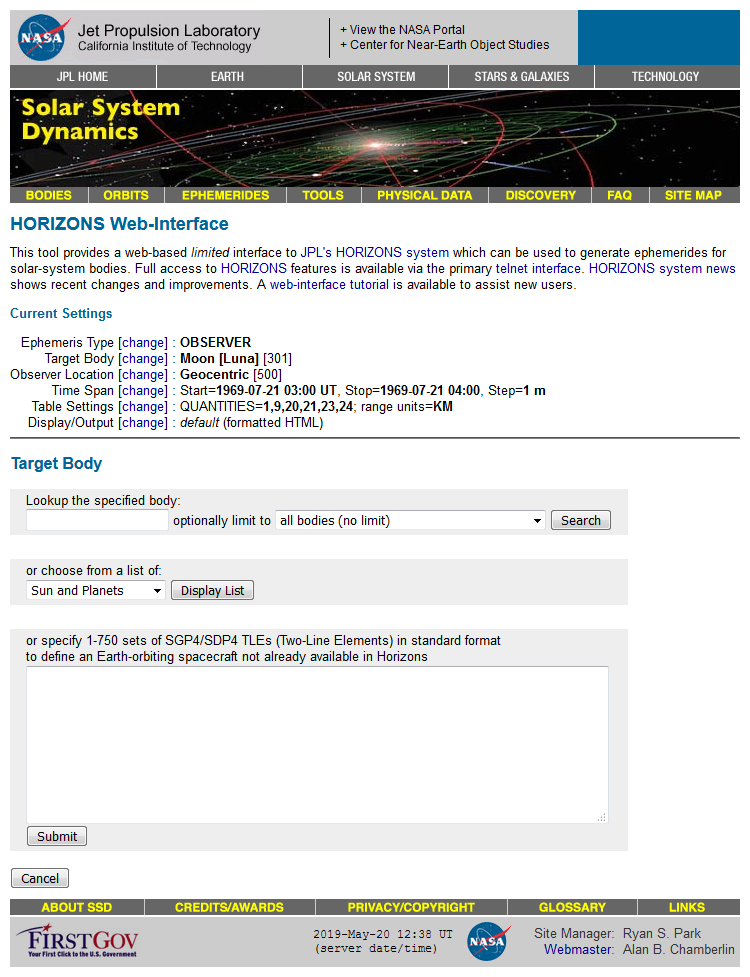}}
\caption{Benutzeroberfläche des JPL/Horizons-Webtools zur Festlegung des Objekts.}
\label{f:jpl2}
\end{figure}

Bei \texttt{Target Body} geben Sie in der neu erscheinenden Eingabemaske (Abb.~\ref{f:jpl2}) unter {\em Lookup the specified body} den Namen \texttt{Luna} ein und klicken auf \texttt{\colorbox{grey}{Submit}}.

\clearpage

Bei \texttt{Time Span} geben Sie bitte als Startzeitpunkt \texttt{1969-07-21 03:30 UT}, als Endzeitpunkt \texttt{1969-07-21 05:00} und als Schrittweite 1 Minute an.

\begin{figure}[!ht]
\centering
\resizebox{0.92\hsize}{!}{\includegraphics{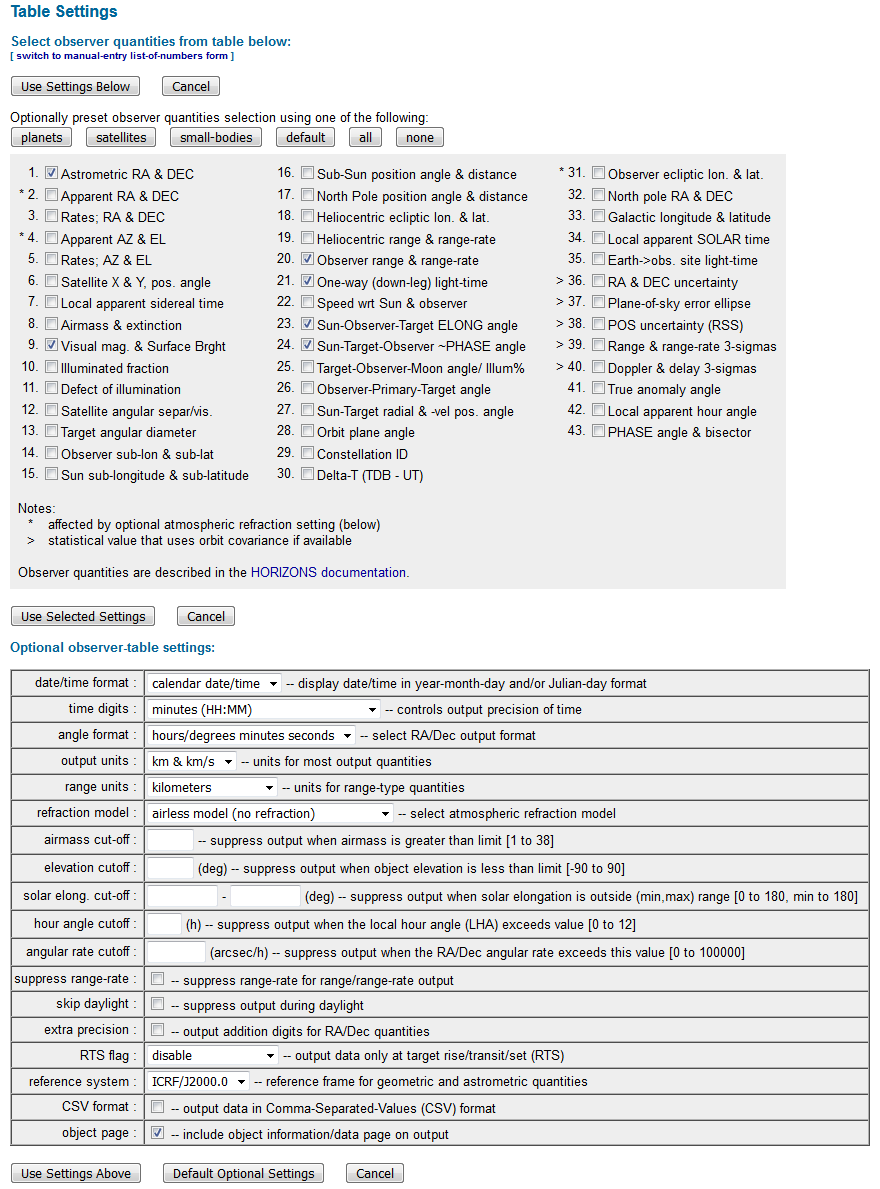}}
\caption{Benutzeroberfläche des JPL/Horizons-Webtools zur Festlegung der zu berechnenden Parameter.}
\label{f:jpl3}
\end{figure}

\medskip
Bei den \texttt{Table Settings} (Abb.~\ref{f:jpl3}) achten Sie darauf, dass der Parameter Nr.~20 ({\em Observer range \& range-rate}) ausgewählt und bei den {\em range units} (untere Tabelle) \texttt{kilometers} eingestellt ist. Klicken Sie dann auf \colorbox{grey}{\texttt{Use settings}}.

\begin{figure}[!ht]
\centering
\resizebox{0.7\hsize}{!}{\includegraphics{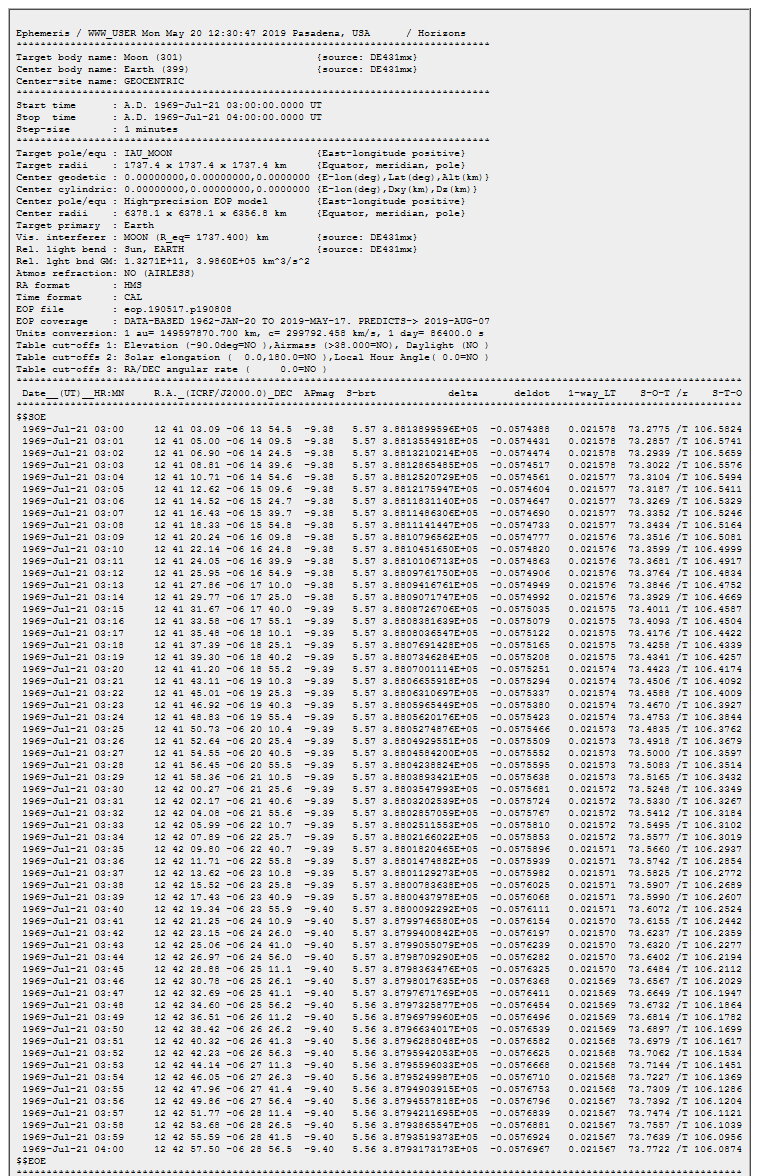}}
\caption{Ausgabe der Bahnparameter des Monds.}
\label{f:jpl4}
\end{figure}

In der Hauptansicht können Sie nun auf \colorbox{grey}{\texttt{Generate Ephemerides}} klicken, und Sie erhalten die entsprechende Liste der Parameter (Abb.~\ref{f:jpl4}). Die Entfernung zwischen Erde und Mond (Parameter: \texttt{delta}) bezieht sich auf die jeweiligen Mittelpunkte der Himmelskörper.

Im Rahmen der Genauigkeit der Analyse reicht es, lediglich einen Wert aus dieser Liste zu entnehmen. Dieser Wert sollte zeitlich etwa in der Mitte der verwendeten Passagen des Funkverkehrs liegen. Ein repräsentativer Wert wäre z.~B. der für 4:00 Uhr.

\vfill
{\em
Hinweis! Um den Schwierigkeitsgrad weiter zu erhöhen, können die Schülerinnen und Schüler anstatt einer mittleren wahren Entfernung für jede Messung der Signallaufzeit die jeweilige wahre Mondentfernung aus dieser Tabelle ableiten. Diese Werte können mit der für jede Messung getrennt zu ermittelnde Mondentfernung verglichen werden. Der Einfluss auf das Endergebnis ist jedoch gering.
}

\clearpage
\subsection*{Diskussion des Ergebnisses}

Vergleichen Sie den von Ihnen ermittelten Wert für den Abstand des Monds von der Erde mit dem wahren Wert. Diskutieren Sie mögliche Einflüsse auf die Differenz der beiden Werte.

\medskip
Berechnen Sie die Differenz. Ermitteln Sie heraus die prozentuale Abweichung des Messergebnisses vom wahren Wert.

\medskip
Berechnen Sie den Betrag der Signallaufzeit, der dieser Abweichung entspricht.

%%%%%%%%%%%%%%%%%%%%%%%%%%%%%%%%%%%%%%%%%%%%%%%%%%%%%%%%%%

\clearpage
\section*{Lösung}

Als Lösungsbeispiel werden aus Tab.~\ref{t:echolist} sieben Passagen analysiert. Die Ergebnisse sind in Tab.~\ref{t:loesung} zusammengefasst. Bei einer erneuten Auswertung können geringe Abweichungen auftreten.

\begin{table}[!ht]
    \caption{Signallaufzeiten für sieben Passagen des Funkverkehrs zwischen Houston und Apollo 11.}
    \label{t:loesung}
    \centering
    \begin{tabular}{lcccccc}
    \hline
   & \multicolumn{2}{c}{Nachricht (mm:ss)} & \multicolumn{2}{c}{Echo (mm:ss)} & \multicolumn{2}{c}{Differenz (s)}\\
Analysierte Textpassage & Beginn & Ende & Beginn & Ende & $\Delta t_1$ & $\Delta t_2$\\
\hline
looks good                           & 34:26,051 & 34:26,489 & 34:28,696 & 34:29,050 & 2,645 & 2,561 \\
\textcolor{grey}{Hou}ston AOS. Over. & 43:33,040 & 43:33,995 & 43:35,663 & 43:36,543 & 2,623 & 2,648 \\
Hou\textcolor{grey}{ston}            & 44:30,074 & 44:30,229 & 44:32,784 & 44:32,879 & 2,710 & 2,650 \\
see the stars                        & 45:18,613 & 45:19,137 & 45:21,255 & 45:21,806 & 2,642 & 2,669 \\
we\underline{t}                      & -- & 60:07,535 & -- & 60:10,186 & -- & 2,651 \\
Houston. Over.                       & 61:14,176 & 61:14,735 & 61:16,859 & 61:17,365 & 2,683 & 2,630 \\
Hou\textcolor{grey}{ston}            & 105:53,887 & -- & 105:56,570 & -- & 2,683 & -- \\
 \hline
    \end{tabular}
\end{table}

\smallskip
Mittelwert: \unit[2,650]{s}

Standardabweichung: \unit[0,037]{s}

SEM: \unit[0,011]{s} (0,4\%)

\medskip
gemessene Entfernung: $\Delta x = c\cdot \frac{\Delta t}{2} = \unit[397163]{km}$ (für ungerundeten Mittelwert)

Standardabweichung: \unit[5587]{km}

SEM: \unit[1613]{km}
\begin{eqnarray*}
\nonumber
    \Delta x_w -  \Delta x_m &=& \unit[-9231]{km} \\[10pt]
\nonumber
    \frac{\Delta x_w -  \Delta x_m}{\Delta x_w} &=& 2,4\%
\end{eqnarray*}

\medskip
Der durch die Analyse der Funksignale ermittelte Wert für die Mondentfernung ist mehr als \unit[9000]{km} zu groß.

\medskip
Da jedoch der Abstand zwischen Sender und Empfänger tatsächlich geringer ist als die Entfernung zwischen den Mittelpunkten von Erde und Mond, ist die Abweichung sogar noch größer. Als Abschätzung können wir annehmen, dass sich die Gesprächspartner ungefähr auf den Oberflächen von Erde und Mond befinden. Bei einer direkten Funkverbindung vermindert sich somit der effektive Abstand der Gesprächspartner um den Radius der Erde plus den Radius des Monds.

\medskip
\begin{equation*}
    \Delta d = r_\textsf{Erde}+r_\textsf{Mond} = \unit[6378]{km}+\unit[1783]{km} = \unit[8161]{km}
\end{equation*}

\medskip
Im Extremfall ist daher die Abweichung in der Mondentfernung um diesen Betrag erhöht, d.~h. insgesamt kann er \unit[17392]{km} betragen, also etwa einen halben Erdumfang. Das entspricht einer Signallaufzeit von \unit[0,058]{s} und einer prozentualen Abweichung von 4,6\% vom theoretischen Wert. Für ein Schulexperiment ist diese Unsicherheit jedoch akzeptabel.

\medskip
Die Abweichung zwischen gemessener und tatsächlicher Mondentfernung lässt sich auf die Ver\-zö\-ge\-rung des Signals während der Übertragung und Weiterleitung auf der Erde zurück führen. Funk- und Kabelstrecken sowie die Verarbeitung der Signale führen zu einer zusätzlichen effektiven Signallaufzeit, die sich in der Messung bemerkbar macht.

%%%%%%%%%%%%%%%%%%%%%%%%%%%%%%%%%%%%%%%%%%%%%%%%%%%%%%%%

% Place Endnotes
\clearpage
 { \parindent 0pt
     \parskip 2ex
    \def\enotesize{\normalsize}
     \theendnotes   }

%%%%%%%%%%%%%%%%%%%%%%%%%%%%%%%%%%%%%%%%%%%%%%%%%%

\clearpage
%BibLatex
\printbibliography

\clearpage
% Danksagung
\section*{Danksagung}
Der Autor bedankt sich bei den Lehrpersonen Matthias Penselin, Florian Seitz, Inge Thiering und Martin Wetz für ihre wertvollen Hinweise, Kommentare und Änderungsvorschläge, die in die Erstellung dieses Materials eingeflossen sind. Weiterer Dank gilt Herrn Dr. Volker Kratzenberg-Annies für seine gewissenhafte Durchsicht.

\medskip
Diese Ausarbeitung basiert auf der Veröffentlichung von \textcite{girlanda_echoes_2009}, in der eine ähnliche Aktivität beschrieben wird.

%\clearpage
\ 
\vfill
\medskip
Diese Unterrichtsmaterialien sind im Rahmen des
Projekts {\em Raum für Bildung} am Haus der Astronomie in Heidelberg entstanden. Weitere Materialien des Projekts finden Sie unter:

\begin{center}
\href{http://www.haus-der-astronomie.de/raum-fuer-bildung}{http://www.haus-der-astronomie.de/raum-fuer-bildung}
und
\href{http://www.dlr.de/next}{http://www.dlr.de/next}
\end{center}

Das Projekt findet in Kooperation mit dem Deutschen Zentrum für Luft- und Raumfahrt statt und wird von der Joachim Herz Stiftung gefördert.\\
\begin{center}
\includegraphics[height=1.5cm]{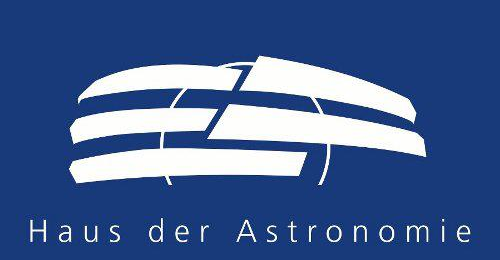}
\hspace*{4em}
\includegraphics[height=1.5cm]{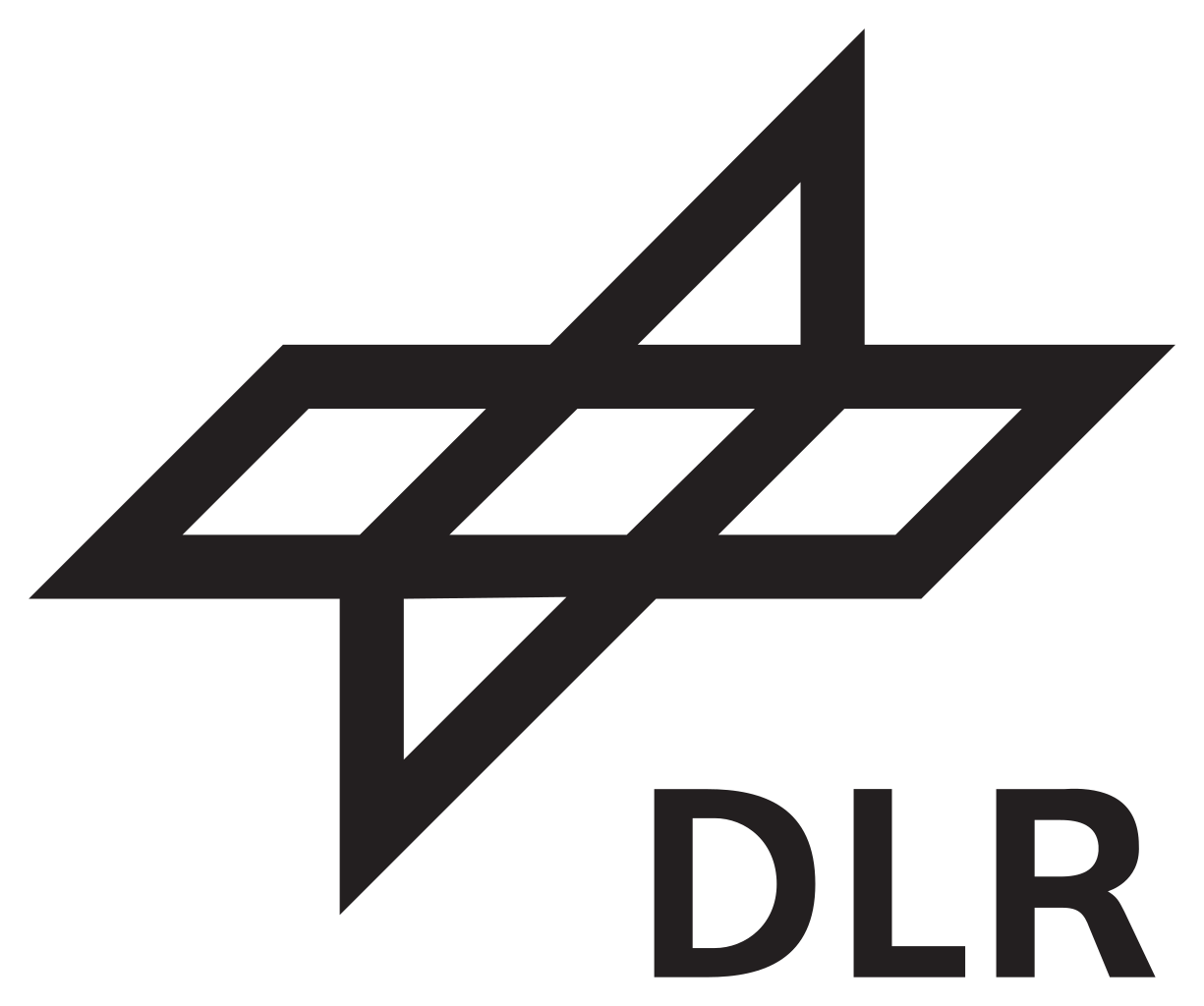}
\hspace*{4em}
\includegraphics[height=1.5cm]{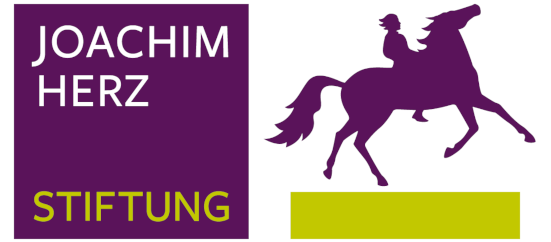}
\end{center}

\end{document}

%% file: header.tex
\usepackage[utf8]{inputenc}
\usepackage[T2A,OT1]{fontenc}
\usepackage[russian,british,ngerman]{babel}

\usepackage{amsmath}

\usepackage{endnotes}

\usepackage[slantedGreek]{cmbright}
\usepackage{units}
\usepackage{ellipsis}

\usepackage{tabularx}
\newcolumntype{L}[1]{>{\raggedright\arraybackslash}p{#1}}
\newcolumntype{C}[1]{>{\centering\arraybackslash}p{#1}}
\newcolumntype{R}[1]{>{\raggedleft\arraybackslash}p{#1}}

\newcommand{\Aud}{\em Audacity}

\usepackage[pdftex]{graphicx}
\usepackage{color}
\definecolor{grey}{gray}{0.7}
\usepackage{hyperref}
\hypersetup{
    bookmarks=true,         % show bookmarks bar?
    unicode=false,          % non-Latin characters in Acrobat’s bookmarks
    pdftoolbar=true,        % show Acrobat’s toolbar?
    pdfmenubar=true,        % show Acrobat’s menu?
    pdffitwindow=false,     % window fit to page when opened
    pdftitle={Wo ist Apollo 11?},    % title
    pdfauthor={Markus Nielbock},     % author
    pdfsubject={Unterrichtsmaterial},   % subject of the document
    pdfcreator={Markus Nielbock},   % creator of the document
    pdfproducer={},  % producer of the document
    pdfnewwindow=true,      % links in new window
    colorlinks=true,       % false: boxed links; true: colored links
    linkcolor=blue,          % color of internal links
    citecolor=blue,        % color of links to bibliography
    filecolor=blue,      % color of file links
    urlcolor=blue           % color of external links
}

\pdftrailerid{} %Remove ID
\usepackage[headsepline,footsepline]{scrlayer-scrpage}
\clearscrheadfoot

\usepackage[bf]{caption}
\newcommand\cyrtext[1]{{\fontencoding{T2A}\selectfont\foreignlanguage{russian}{#1}}}

\newcommand\leftmarkline[1]{%
  \parbox[c][\layerheight][b]{\layerwidth}{%
    \hspace*{3.5mm}\rule{#1}{.2mm}%
}}
\DeclareNewLayer[{
  background,
  innermargin,
  oddpage,% bei zweiseitigen Dokumenten nur auf den ungeraden Seiten
  height=.34\paperheight,
  contents={\leftmarkline{2mm}}
}]{oberefaltmarke}
\DeclareNewLayer[{
  clone=oberefaltmarke,
  height=.67\paperheight
}]{unterefaltmarke}
\DeclareNewLayer[{
  clone=oberefaltmarke,
  height=.5\paperheight,
  contents={\leftmarkline{4mm}}
}]{lochermarke}
\AddLayersToPageStyle{@everystyle@}{lochermarke}

\setkomafont{pagefoot}{\normalfont\footnotesize}
\setkomafont{caption}{\normalfont\footnotesize}

\usepackage[babel,german=quotes]{csquotes} %deutsches Anführungszeichen

\usepackage[backend=biber,style=authoryear-icomp]{biblatex}
\newcommand{\myproject}{\Large Unterrichtsmaterialien zum\\ 50-jährigen Jubiläum von Apollo 11}
\newcommand{\mytitle}{Wo ist Apollo 11?}
\newcommand{\myshorttitle}{Wo ist Apollo 11?}
\newcommand{\myclasses}{Klassen 10-13}
\newcommand{\myauthor}{Markus Nielbock}

\rehead*{In Kooperation mit\\ \includegraphics[height=11mm]{DLR_Logo_svg.png}}
\rohead*{In Kooperation mit\\ \includegraphics[height=11mm]{DLR_Logo_svg.png}}
\lehead*{\includegraphics[height=15mm]{HdA_500x260.png}}
\lohead*{\includegraphics[height=15mm]{HdA_500x260.png}}
\ofoot*{Seite \pagemark}
\ifoot*{Handreichung: \myshorttitle}

%% file: titlepage.tex
%%%%%%%%%%%%%%%%%%%%%%%%% Titlepage %%%%%%%%%%%%%%%%%%%%%%%%%%%%%%%%%%%%
\begin{titlepage}
\thispagestyle{scrheadings}

\begin{center}
{\large\bfseries
Handreichung für Lehrpersonen:

\bigskip
\LARGE
\mytitle
}

\bigskip
{\large\bfseries
\myclasses
}

 \bigskip
{\bfseries
\myauthor
}

\bigskip
27.~Juni 2019
\end{center}

\section*{Zusammenfassung}
Die Schülerinnen und Schüler analysieren Audiodateien des Funkkontakts zwischen der NASA-Bodenstation in Houston, Texas und der Crew von Apollo 11 während der Mondlandung im Jahr 1969. Durch Echos in der Funkübertragung ermitteln sie die Signallaufzeit und somit die Entfernung zwischen Erde und Mond.

\section*{Lernziele}
Die Schülerinnen und Schüler
\begin{itemize}
\item machen sich mit dem Apollo-Programm vertraut,
\item werten den Original-Funkverkehr des Apollo 11-Programms aus,
\item errechnen die Signallaufzeit der Funkverbindung,
\item vergleichen die ermittelte Mondentfernung mit dem tatsächlichen Wert.
\end{itemize}

\section*{Materialien}
\begin{itemize}
\item Arbeitsblätter (erhältlich unter: \url{http://www.haus-der-astronomie.de/raum-fuer-bildung})
\item Computer (PC, Laptop) mit Kopfhörer (je einer pro Gruppe), ggf. mit Internetzugang
\item Software {\Aud} (\url{https://www.audacity.de/})
\item Audiodatei \texttt{175-AAA.mp3} (\url{https://archive.org/download/Apollo11Audio/175-AAA.mp3})
\item optional: Tabellenkalkulations-Software (Excel) für statistische Auswertung
\item Taschenrechner
\item Stift und Papier für Notizen
\end{itemize}

\section*{Stichworte}
Mond, Apollo 11, Lichtgeschwindigkeit, Signallaufzeit, Ranging, Entfernung

\section*{Dauer}
90 bis 150 Minuten (je nach Auswahl der zu analysierenden Passagen)

\end{titlepage}

%% file: RaumBildung.bib
@online{dupas_gab_1994,
	title = {Gab es einen Wettlauf zum Mond?},
	url = {https://www.spektrum.de/magazin/gab-es-einen-wettlauf-zum-mond/821729},
	titleaddon = {Spektrum.de},
	author = {Dupas, Alain and Logsdon, John M.},
	urldate = {2019-04-03},
	date = {1994-08-01},
	langid = {german}
}

@book{orloff_apollo_2000,
	location = {Washington, D.C.},
	title = {Apollo by the numbers: a statistical reference},
	isbn = {0-16-050631-X},
	url = {https://ntrs.nasa.gov/archive/nasa/casi.ntrs.nasa.gov/20010008244.pdf},
	shorttitle = {Apollo by the numbers},
	number = {{NASA} {SP}-2000-4029},
	publisher = {National Aeronautics and Space Administration},
	author = {Orloff, Richard W},
	urldate = {2019-04-05},
	date = {2000},
	langid = {american},
	note = {{OCLC}: 651892628}
}

@article{girlanda_echoes_2009,
	title = {Echoes from the Moon},
	volume = {77},
	issn = {0002-9505, 1943-2909},
	url = {http://aapt.scitation.org/doi/10.1119/1.3098261},
	doi = {10.1119/1.3098261},
	pages = {854--857},
	number = {9},
	journaltitle = {American Journal of Physics},
	author = {Girlanda, Luca},
	urldate = {2019-04-17},
	date = {2009-09},
	langid = {english}
}

@inproceedings{yaplee_lunar_1959,
	location = {Redwood City, California, USA},
	title = {A lunar radar study at 10-cm wavelength},
	volume = {9},
	url = {http://www.journals.cambridge.org/abstract_S0074180900050506},
	doi = {10.1017/S0074180900050506},
	series = {{IAU} Symposium},
	eventtitle = {Paris Symposium on Radio Astronomy},
	pages = {19--28},
	booktitle = {Proceedings of the International Astronomical Union},
	publisher = {Stanford University Press},
	author = {Yaplee, B. S. and Roman, Nancy G. and Craig, K. J. and Scanlan, T. F.},
	editor = {Bracewell, Ronald N.},
	urldate = {2019-04-17},
	date = {1959},
	langid = {english}
}

@article{soffel_lasermessungen_1997,
	title = {Lasermessungen der Monddistanz},
	volume = {7/1997},
	url = {https://www.spektrum.de/inhaltsverzeichnis/juli-1997/1144985},
	pages = {646--651},
	journaltitle = {Sterne und Weltraum},
	shortjournal = {{SuW}},
	author = {Soffel, Michael H. and Müller, Jürgen},
	urldate = {2019-04-17},
	date = {1997}
}

@article{yaplee_radar_1958,
	title = {Radar Echoes from the Moon at a Wavelength of 10 {CM}},
	volume = {46},
	issn = {0096-8390},
	url = {http://ieeexplore.ieee.org/document/4065257/},
	doi = {10.1109/JRPROC.1958.286790},
	pages = {293--297},
	number = {1},
	journaltitle = {Proceedings of the {IRE}},
	author = {Yaplee, B. and Bruton, R. and Craig, K. and Roman, N.},
	urldate = {2019-04-17},
	date = {1958-01}
}

@report{corliss_histories_1974,
	title = {Histories of the Space Tracking and Data Acquisition Network ({STADAN}), the Manned Space Flight Network ({MSFN}), and the {NASA} Communications Network ({NASCOM})},
	rights = {Unclassified; Publicly available; Unlimited},
	url = {https://ntrs.nasa.gov/search.jsp?R=19750002909},
	pages = {360},
	number = {{NASA} {CR}-140390},
	institution = {{NASA}},
	type = {Technical Report},
	author = {Corliss, William R.},
	urldate = {2019-05-06},
	date = {1974-06-01},
	langid = {american}
}

@online{sapienza_nasa_2009,
	title = {{NASA} - 1969 Apollo 11 News Release},
	url = {https://www.nasa.gov/centers/glenn/about/history/apollo_press_release.html},
	type = {Feature Articles, Feature},
	author = {Sapienza, Jennifer},
	editora = {Kennard, Emily and Dunbar, Brian and Zona, Kathleen},
	editoratype = {collaborator},
	urldate = {2019-04-24},
	date = {2009},
	langid = {english}
}

@report{graham_manned_1970,
	location = {Greenbelt, Maryland, {USA}},
	title = {Postmission Report on the AS-506 (Apollo 11) Mission: Manned Space Flight Network},
	url = {https://www.hq.nasa.gov/alsj/a11/AS-506-PMR.pdf},
	shorttitle = {Apollo 11 {MSFN} Postmission Report},
	institution = {Goddard Space Flight Center},
	author = {Graham, D. J. and Roberts, Carl O. and Wood, H. William},
	urldate = {2019-05-06},
	date = {1970-02},
	langid = {american}
}
